%% file: Harding_MSPemission-update.tex
\documentclass[graybox]{svmult}

\usepackage{type1cm}        
\usepackage{makeidx}         
\usepackage{graphicx}        
\usepackage{multicol}        
\usepackage[bottom]{footmisc}

\usepackage{newtxtext}       %
\usepackage{newtxmath}       

\makeindex             

\def\lsim{\lower 2pt \hbox{$\, \buildrel {\scriptstyle <}\over
         {\scriptstyle \sim}\,$}}
         
\begin{document}

\title*{The emission physics of millisecond pulsars}

\author{Alice K. Harding}
\institute{Alice K. Harding \at Theoretical Division, Los Alamos National Laboratory, Los Alamos, NM 87545 \email{ahardingx@yahoo.com}}
%
%
\maketitle

\abstract*{Understanding the physics of rotation-powered millisecond pulsars (MSPs) presents a number of challenges compared to that of the non-recycled pulsar population.  Even though their fast rotation rates can produce high spin-down power and accelerating electric fields, their relatively low surface magnetic fields make the production of electron-positron pairs required for radio emission difficult.  The Fermi Gamma-Ray Space Telescope has discovered pulsed $\gamma$-rays from a large fraction of the MSP population that have light curves surprisingly similar to those of young pulsars.  However, their very compact magnetospheres enable magnetic fields at the light cylinder that are comparable to those of the most energetic pulsars.  This fact and recent global magnetosphere models showing that particle acceleration takes place near and beyond the light cylinder, now makes the $\gamma$-rays from MSPs plausible.  The large increase in binary systems harboring MSPs has revitalized the study of shock acceleration and high-energy emission in such systems, with many showing orbitally-modulated X-rays.  This Chapter will review the history and our current studies of the mechanisms for multiwavelength emission from MSPs.}

\abstract{Understanding the physics of rotation-powered millisecond pulsars (MSPs) presents a number of challenges compared to that of the non-recycled pulsar population.  Even though their fast rotation rates can produce high spin-down power and accelerating electric fields, their relatively low surface magnetic fields make the production of electron-positron pairs required for radio emission difficult.  The Fermi Gamma-Ray Space Telescope has discovered pulsed $\gamma$-rays from a large fraction of the MSP population that have light curves surprisingly similar to those of young pulsars.  However, their very compact magnetospheres enable magnetic fields at the light cylinder that are comparable to those of the most energetic pulsars.  This fact and recent global magnetosphere models showing that particle acceleration takes place near and beyond the light cylinder, now makes the $\gamma$-rays from MSPs plausible.  The large increase in binary systems harboring MSPs has revitalized the study of shock acceleration and high-energy emission  in such systems, with many showing orbitally-modulated X-rays.  This Chapter will review the history and our current studies of the mechanisms for multiwavelength emission from MSPs.}

\section{Introduction}
\label{sec:intro}

Rotation-powered pulsars display emission over a broad range of wavelengths from radio to TeV $\gamma$-rays.  Millisecond pulsars (MSPs) have broadband spectra that are similar to those of non-recycled (normal) pulsars, with coherent radio emission components, incoherent UV to X-ray thermal and/or non-thermal components and non-thermal $\gamma$-ray components that extend to GeV energies.  Differences in emission between normal pulsars and MSPs are present but more subtle.  MSPs tend to have on average steeper radio spectra and radio pulse profiles that are wider and more complex (see Chapter 1 for more details), hotter thermal X-ray spectra and harder $\gamma$-ray spectra (see Chapter 2), and to date no MSPs have detected emission at energies greater than 10 GeV.  By contrast, several young pulsars have detected pulsed emission above 50 GeV, including the Crab pulsar with pulsations up to 2 TeV (Ansoldi et al. 2016), the Vela pulsar with pulsations up to 7 TeV (Djannati-Atai et al. 2017), the Geminga pulsar with pulsations above 20 GeV (Lopez et al. 2019) and PSR B1706-44 with pulsations up to 70 GeV (Spir-Jacob et al. 2019).  Despite these differences, the physics and mechanisms of MSP emission is much the same as for normal pulsars.  Due to their very rapid rotation rates (up to several hundred Hz) and despite their much lower surface magnetic fields ($\sim 10^8$ G), MSPs have higher spin-down power on average than normal pulsars.  There are, however, challenges to emission modeling that are specific to MSPs.  The low surface magnetic fields greatly hinder magnetic one-photon pair production in a dipole field that is needed for the operation of the pair cascades that are thought to be responsible for both the creation of plasma for the magnetosphere and for radio emission.  This problem, as well as recent observational results,  strongly suggest that non-dipolar magnetic fields are present near the surface of MSPs.  Therefore, complex surface field structure may be a hallmark and a driver of MSP emission physics.  It may also be a direct consequence of their history of spin-up by accretion in Low Mass X-ray Binaries (LMXB) (See Chapter 9).

Consequently, this Chapter will begin with a review of the structure and physics of the global pulsar magnetosphere, as we presently understand it, pointing out the yet unsolved problems.  We will then review models for the broadband emission, with specific predictions for MSPs.  The physics of pair cascades and their operation in MSPs, as well as the implications for radio death lines and magnetic field structure, are discussed in Section \ref{sec:pairs}.  In Section \ref{sec:PC}, the heating of the neutron star surface by energetic particles from the pair cascades, which produce the observed thermal X-ray components, will be reviewed.  The large number of $\gamma$-ray MSPs discovered by the {\it Fermi Large Area Gamma-Ray Telescope} has led to a factor of seven increase in the number of exotic binary systems known as Black Widows and Redbacks, whose MSPs are heating and ablating their companions with their pulsar winds, producing intra-binary shocks that are accelerating particles.  Section \ref{sec:bin} will review the high-energy emission from these systems.  Finally, Section \ref{sec:probs} will discuss the outstanding problems in modeling MSP emission.

\section{Global Magnetosphere Models}
\label{sec:mag}

Since the emission physics of pulsars is closely tied to the geometry of the magnetic and electric fields, it is important to first discuss the global solutions for the electrodynamics of a rotating dipole field and how these have evolved with increasingly better computational power.  

\subsection{Vacuum retarded dipole}
\label{sec:VRD}

At the time pulsars were first discovered in 1967, the only global magnetosphere solution that existed was the vacuum retarded dipole (VRD) (Deutsch 1955) that describes the electric and magnetic fields of star rotating at rate $\Omega$ with a magnetic dipole field in vacuum.  The analytic VRD solution showed that a very strong quadrupole electric field is induced near the star with a large component parallel to the magnetic field.  This electric field falls off as $1/r^4$, much faster than the magnetic dipole field, $1/r^3$.  At large distance from the star, $r \sim R_{\rm LC} = c/\Omega$, where the corotation velocity reaches c the displacement current distorts the magnetic field, causing it to sweep back opposite to the direction of rotation.  Well beyond the light cylinder, the solution becomes an electromagnetic wave as imposed by the far-field boundary condition.

Although the VRD magnetic field structure was a more realistic alternative than the static vacuum dipole and was used over several decades for pulsar emission modeling (see Section \ref{sec:emiss}), it is completely lacking some essential elements needed for a physical description of a pulsar magnetosphere, namely charges and currents.  Very early in the pulsar modeling game, Goldreich \& Julian (1969) noted that a magnetized neutron star  cannot be surrounded by a vacuum since the strong parallel electric fields, $E_\parallel$, about the surface would pull charges from the star  to fill the magnetosphere.   This concept of a magnetosphere filled with plasma however, was not easy to fully model.  Early N-body simulations (Krauss-Polsdorff \& Michel 1985),  allowing either sign of charge to be pulled from the neutron star and accelerated by the vacuum $E_\parallel$ for an aligned rotator failed to fill the magnetosphere with charge.  Rather, it showed that two separated regions of static and opposite charge form above the polar caps (dome) and along the equator (torus) with no currents, producing a dead pulsar (electrosphere) solution.  However, with the advent of global kinetic plasma simulations (see Section \ref{sec:PIC}) it has been shown that diocotron instabilities (Petri et al. 2002) as well as the production of electron-position pairs above the neutron star surface prevents a stable electrosphere solution.

\subsection{Force-free and dissipative models}
\label{Sec:FF}

The Force-Free (FF) magnetosphere is the solution to Maxwell's Equations for a rotating star with a dipole field, ignoring plasma pressure and requiring that the electric field parallel to the magnetic field is zero everywhere (${\bf E \cdot B} = 0$).  In this special case (force-free and ideal MHD), the charge density in the magnetosphere is equal to the Goldreich-Julian charge (Goldreich \& Julian 1969), $\rho_{\rm GJ} = -{\bf \Omega \cdot B}/2\pi c$, where $\Omega$ and $B$ are the pulsar rotation rate and magnetic field, which is the density required to locally screen the $E_\parallel$. The equation describing the fields and currents of an aligned FF magnetosphere, the `Pulsar Equation' introduced by Michel (1973), was first solved numerically by Contopoulos et al. (1999).   The current density distribution across the polar cap in this solution shows that the current flows out along field lines near the magnetic poles and returns to the star along a current sheet (where magnetic field lines of opposite polarity merge) and separatrix (along the last open field line that separates the open and closed magnetosphere, those field lines that close within $R_{\rm LC}$) (see also Timokhin  2006).  Numerical solutions for oblique FF magnetospheres were derived by solving the time-dependent Maxwell's Equations (Spitkovsky 2006, Kalapotharakos \& Contopoulos 2009), showing that the return current region becomes more distributed and axisymmetric with increasing inclination angle.  FF magnetospheres have polar caps that are larger that those of vacuum magnetospheres and shifted backward toward the trailing edge (Bai \& Spitkovsky 2010, Harding et al. 2011).

The FF models however still do not describe real pulsars since there is no $E_\parallel$ acceleration and no radiation.  If the FF condition ${\bf E\cdot B} = 0$ is relaxed, dissipative magnetosphere solutions can be found for different values of a macroscopic conductivity $\sigma$ (Kalapotharakos et al. 2012a, Li et al. 2012, Cao et al. 2016).  These solutions span the range between vacuum and FF magnetospheres and showed self-consistent regions of $E_\parallel$ that exist along the separatrix and current sheet.   However, the dissipative models are not completely self-consistent since the microphysics that creates the $\sigma$ distribution is not specified.  

\subsection{Kinetic models}
\label{sec:PIC}

\begin{figure}[t]
\hskip -0.6cm
\includegraphics[width=130mm]{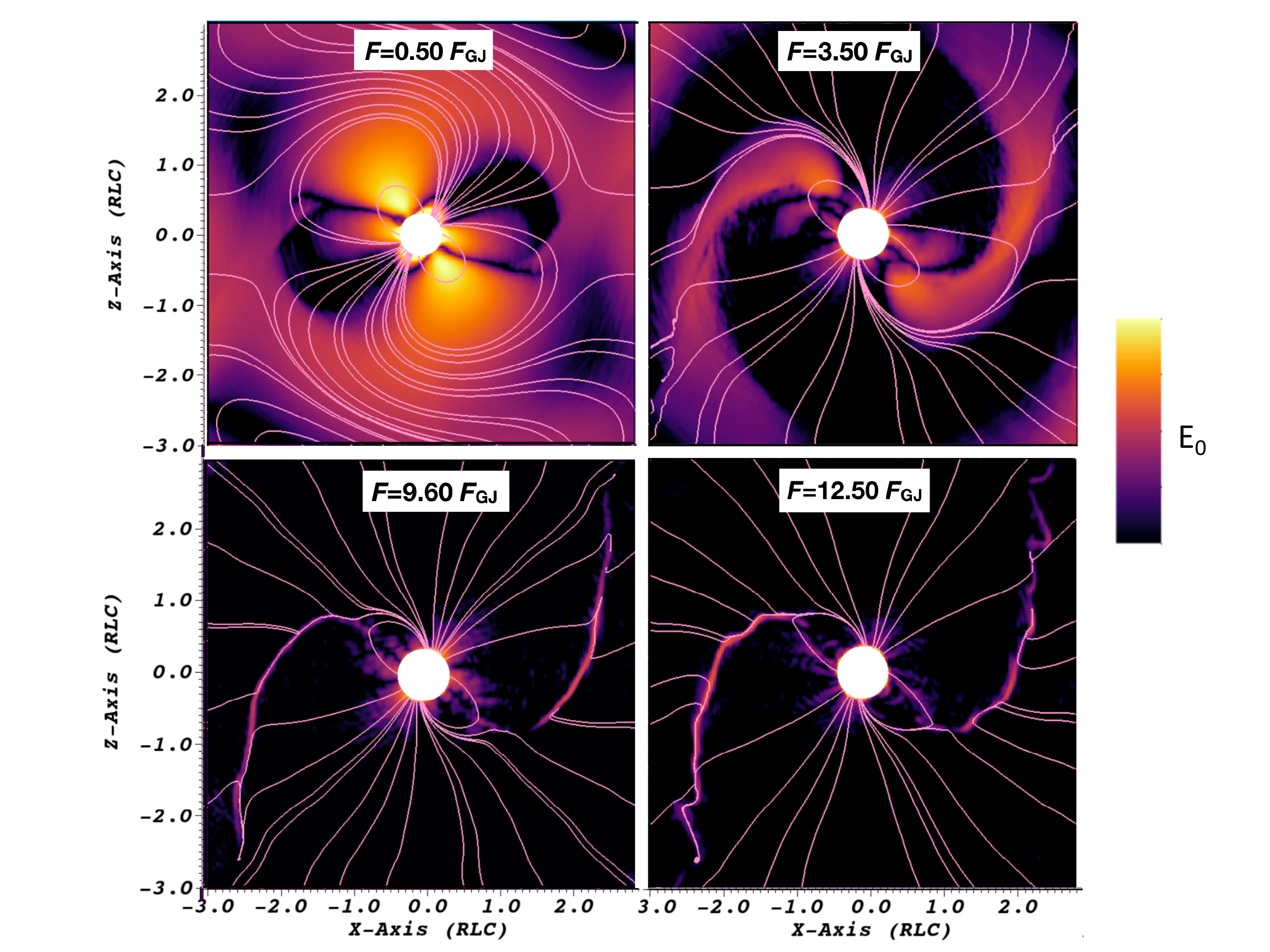}
\caption{3D kinetic simulations of a global pulsar magnetosphere with $\alpha = 45^\circ$ showing the magnetic field (pink lines) and accelerating electric field $E_0$ distribution (color scale) in the poloidal plane of the dipole and spin axis, for different pair injection rates in units of the Goldreich-Julian rate.  From Brambilla et al. (2018).}
\label{fig:Epar}       
\end{figure}

To fully compute the self-consistent feedback between particle motions and fields, kinetic plasma simulations are needed.  Particle-in-Cell (PIC) models provide a method to track the feedback between particles and fields by solving the time-dependent Maxwell's Equations and the particle equations of motion simultaneously (actually in alternating time steps).  The first PIC simulations of a pulsar magnetosphere were performed by Philippov \& Spitkovsky (2014), using a 3D Cartesian grid and injecting pair plasma ($e^+-e^-$ pairs) throughout the computational domain and by Chen \& Beloborodov (2014) using a 2D spherical grid.  Cerutti et al. (2015) using a spherical 2D PIC code showed that injecting enough pair plasma only above the neutron star surface could produce a near-FF solution for an aligned rotator.  Simulations requiring arbitrary thresholds on particle energies for pair injection found that pair production must occur in the current sheet as well as at the polar caps to create a FF solution (Chen \& Belodorodov 2014, Philippov et al. 2015).  Using a 3D Cartesian PIC code, Kalapotharakos et al. (2018) found that if a larger rate of pairs is injected, everywhere or just from the neutron star surface (Brambilla et al. 2018), FF magnetospheres can be formed at all inclination angles without the need for pair production in the current sheet.  For increasing injection rates, the $E_\parallel$ is screened over a larger part of the magnetosphere and confined more to the current sheet, where the highest energy particles are found (see Figure \ref{fig:Epar}).

\section{Multiwavelength Emission Models}
\label{sec:emiss}

\begin{figure}[t]
\hskip -1.0cm
\includegraphics[width=130mm]{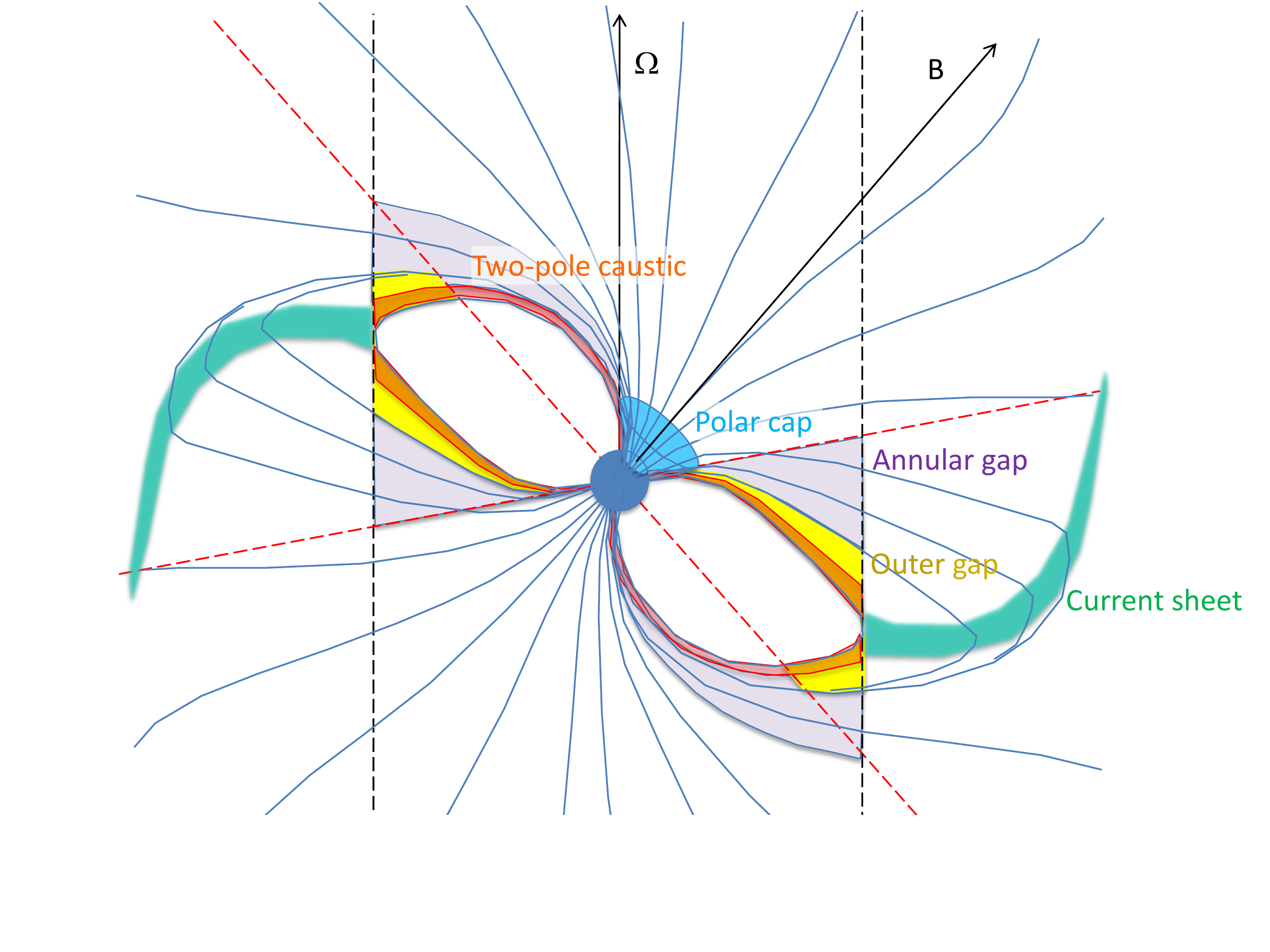}
\vskip -1.0cm
\caption{Location of polar cap/radio (blue) and high-energy emission in the outer gap (yellow), two-pole caustic (red), annular gap (purple) and current sheet (green) models in the meridional plane containing the spin and magnetic axes of a near force-free magnetosphere.  The dashed black lines denote the light cylinder and the dotted red lines show projections of the null-charge surface.}
\label{fig:PSRmods}       
\end{figure}

It is clear that pulsar emission occurs at a variety of different sites throughout the magnetosphere across wavelength bands (see Figure \ref{fig:PSRmods}).  From the morphology of pulse profiles and their polarization, the radio emission seems to be originating on open field lines above the polar caps.   Many normal pulsars show the ``S"-shaped swing of polarization position angle predicted by the Rotating Vector Model (Radhakrishnan \& Cooke 1969).  MSP radio emission is generally more complicated than that of normal pulsars, with wider, more complex profiles and polarization patterns that are harder to interpret.  By contrast, the high-energy emission from MSPs (see Chapter 2) is more similar to that of normal pulsars.  Thanks to the many detections of pulsed $\gamma$-ray emission from both normal and MSPs by {\it Fermi}  (Abdo et al. 2013), it is now clear that most of the high-energy emission from pulsars comes from the outer magnetosphere.  As discussed in Section \ref{sec:mag}, global magnetosphere models have shown that most of the particle acceleration and high-energy emission occurs near the current sheet outside the light cylinder.  In the global models, the structure of the fields, currents and current sheets scale with the light cylinder independent of the pulsar period or surface magnetic field strength.  Therefore, similarity of the high-energy emission in normal pulsars and MSPs is to be expected.  However, even at high energies, there are notable differences both observationally and theoretically.  In the X-ray domain, two groups of MSPs with different emission characteristics have often been discussed, with one group being more energetic with spectra dominated by non-thermal emission and narrow pulse profiles, and the other group being less energetic with spectra dominated by thermal emission and broad, sinusoidal-like pulse profiles (Becker 2009).  At GeV energies, there are three classes of MSPs with distinct $\gamma$-ray profile characteristics: Class I show narrow double-peaked profiles where the first peak trails the radio peak by one or two tenths of a period, behavior very similar to that of normal $\gamma$-ray pulsars;  ; Class II have phase-aligned $\gamma$-ray and radio profiles (Guillemot et al. 2012), like the Crab pulsar and PSR B0540-69; Class III  have broader, non-standard $\gamma$-ray profiles with the first peak leading the radio peak, a behavior unique to MSPs.  Class I  MSP $\gamma$-ray profiles tend to have larger phase lags with the radio peaks than do normal pulsars.  

\subsection{Polar cap models}

Emission models for MSPs have paralleled those of normal pulsars, with the early models placing the high-energy emission at the polar caps followed by models placing the emission near (but inside) the light cylinder and present-day models having emission near the current sheet.  Because of their short periods and larger polar caps, MSPs have voltage drops above the  polar caps that are similar to some young pulsars.  Consequently, polar cap models for normal pulsars (Daugherty \& Harding 1996) were adopted for MSP high-energy emission (Bulik et al. 2000,Venter et al. 2005) and predicted detectable fluxes of curvature radiation (CR) and inverse-Compton scattering (ICS) emission from primary accelerated particles scattering thermal X-rays from the neutron star polar cap.  However, because MSPs have very low surface magnetic fields, particles must be accelerated to higher energies to emit more energetic photons in order to pair produce.  Also, since the field line radius of curvature is smaller due to the larger polar caps, the CR energy loss rate is higher and the particles may become radiation-reaction limited before producing pair-creating photons (Luo et al. 2000).  In the extreme case, which may include most MSPs as well as older normal pulsars, the voltage drop will not be high enough to produce enough electron-positron pairs to screen the fields near the polar caps, allowing acceleration of particles and radiation to extend to high altitude along the open field lines (Muslimov \& Harding 2004a).  In such ``pair-starved polar cap" (PSPC) models, the high-energy emission could be visible over a much larger fraction of the sky compared to polar cap emission that is limited to a very small emission cone.  Harding et al. (2005) applied the PSPC regime to model high-energy emission from MSPs, proposing that the primary CR spectrum would extend above 10 GeV if the observer line of sight crossed near the magnetic poles.  They also predicted that {\it Fermi} would detect the cumulative emission from the many MSPs that inhabit globular clusters. 

\subsection{Outer gap models}

Vacuum gaps in the outer magnetosphere can develop in a charge-separated magnetosphere between the null charge line where the Goldreich-Julian charge, $\rho_{\rm GJ}$, changes sign and the light cylinder, if particles of one sign flow out along open field lines but there are no charges below to replace them (Cheng et al. 1986).   Outer gaps also require that there be no polar cap pair cascades that send a super-Goldreich-Julian particle flux into the gap.  
The gap accelerates primary charges to produce CR, but near the light cylinder pairs are created by the photon-photon process, using thermal X-rays from the neutron star surface, since the field is too low for magnetic pair production.  The 3D geometry of the outer gap was modeled by Romani \& Yadigaroglu (1995), using the Deutsch (1955) retarded vacuum dipole magnetic field and emission up to but inside the light cylinder, producing $\gamma$-ray light curves with one or two peaks from the same magnetic hemisphere.  
 
High-energy emission from outer gaps of MSPs has been explored by Zhang \& Cheng (2003) and Zhang et al. (2007) (see also Takata et al. 2012).  In outer gap models for normal pulsars, the polar cap heating from accelerated particles flowing down from the gap give a surface temperature much higher than observed.  The solution to this problem is to reflect the hot thermal emission back to a larger fraction of the neutron star surface with resonant Compton scattering by particles near the star created in pair cascades (Zhang et al. 1997a).   However, for MSPs with very low surface fields, the pair cascades from downward flowing gap-accelerated particles do not occur for dipole fields so that multipole fields near the neutron star surface are required to create enough pairs.  The predicted spectra consist of two thermal X-ray components (one hotter from a heated polar cap and one cooler from the emission reflected to most of the neutron star), a non-thermal downward-going X-ray SR component from near-surface pairs and an upward-going high energy $\gamma$-ray synchro-curvature  component from the outer gap.  This model thus predicts that the phase shift between the non-thermal X-ray pulse and $\gamma$-ray pulse will be large, which is not consistent with MSPs such as PSR B1821-24 and PSR J0218+4232 whose non-thermal X-ray and $\gamma$-ray peaks are in phase.  
Using a similar outer gap model, Jiang et al. (2014) were able to fit a number of phase-averaged $\gamma$-ray spectra of MSPs measured by {\it Fermi}.
Lyutikov (2013) proposed a model where the high-energy (GeV) emission is cyclotron self-Compton radiation by a spectrum of particles accelerated and counter-streaming in the outer gaps.  This model can produce the observed Crab spectrum from UV to VHE $\gamma$-rays.

\subsection{Slot gap and annular gap models}

In polar cap models, the Poisson Equation solution for $E_\parallel$ varies over the polar cap, decreasing to zero at the last open field line that borders the force-free fully conducting closed field region.  The slot gap (SG) is a narrow set of field lines near this boundary in which particles accelerate more slowly and never radiate photons with high-enough energy to produce pairs before the magnetic field drops too low.  In the SG, the $E_\parallel$ is never screened by pairs and particle acceleration and radiation continue up to high altitude (Arons 1983).  High-energy emission can thus be produced over a large range of radii from the neutron star surface to near the light cylinder (Muslimov \& Harding 2004b).  Along trailing field lines, the observed phase differences of radiation due to the curved field is canceled by aberration and light travel time delays, forming caustics so that all the radiation arrives to the observer in phase.  In such a two-pole caustic model (TPC, Dyks \& Rudak 2003) caustic patterns from both magnetic poles form one or two peaks with variable spacing in the light curves.  The slot gap produces light curves like those of the TPC model, with varied separation of peaks that have phase lags with the radio peaks (Harding et al. 2008), similar to those of {\it Fermi} pulsars (see Section \ref{sec:LC}).  

The annular gap model is geometrically similar to the slot gap with two distinct regions of the polar cap: the annular field lines are those that cross the null charge surface within the light cylinder and are located at the outer edge of the polar cap while the core field lines are those interior to the annular gap (Qiao et al. 2004, 2007).  Unlike the outer gap, the annular gap model assumes that the magnetosphere is not charge-separated but consists of a quasi-neutral pair plasma.  In the core region, the $E_\parallel$ is screened at a low altitude by pairs since the local charge density $\rho$ has the same sign as $\rho_{\rm GJ}$, but in the annular gap the pairs do not screen the $E_\parallel$ since $\rho$ has the opposite sign as $\rho_{\rm GJ}$ and acceleration can continue to high altitude.  The caustics of the $\gamma$-ray light curves from annular gaps are similar to those of the slot gap and can also fit many of the {\it Fermi} pulsar light curves (Du et al. 2010) and in particular several MSPs (Du et al. 2013).

\subsection{Current sheet models}
\label{sec:CS}

Pulsed high-energy emission from the current sheet was first proposed by Kirk et al. (2002) based on the striped pulsar wind discussed by Coroniti (1990).  The emission was assumed to come from a region at 10 to 100 $R_{\rm LC}$ where the field energy of the wind is at least partly converted to particle energy.   The particles then radiate Doppler-boosted SR and an observer will see a pulse of emission each time the current sheet crosses the light-of-sight.  With the emissivity assumed to be uniform throughout the current sheet, the radio lags are larger than those seen in {\it Fermi} pulsar light curves.  Later versions of the striped-wind model (Petri 2011, 2012) using the geometric wind solution of Bogodalov (1999) placed the emission closer to the $R_{\rm LC}$, within 50 $R_{\rm LC}$, and modeled the predicted polarization signatures that could match the observed optical polarization of the Crab pulsar.  

Dissipative global MHD models show that the regions of acceleration do lie mostly outside the light cylinder near the current sheet.  
Using global models with infinite conductivity $\sigma$ (FF) inside the light cylinder and finite $\sigma$ outside the light cylinder and assuming that the accelerated particles radiate CR at GeV energies, Kalapotharakos et al. (2014) found that the predicted particle acceleration and radiation patterns matched the characteristics of the {\it Fermi} pulsars.  Phase-resolved spectra in dissipative magnetospheres (Brambilla et al. 2015) are able to account for the variation of the spectral index and cutoff energy as a function of pulse phase seen for bright {\it Fermi} pulsars (DeCesar 2013).  It was also seen  that $\sigma$ should increase with spin-down power $\dot E$.  Since $\sigma$ in these models is physically tied to the density of pairs in different parts of the magnetosphere, it is expected that pulsars with higher $\dot E$ create a larger number of pairs.  

Harding \& Kalapotharakos (2015) used the magnetic field structure of FF magnetospheres to model broad-band radiation from young pulsars and MSPs, calculating trajectories of particles along field lines from the neutron star surface to $2 R_{\rm LC}$.  CR, SR and synchrotron-self Compton (SSC) emission was produced by both primary particles accelerating in an assumed $E_\parallel$ and electron-positron pairs produced in polar cap pair cascades.  The computed spectra show pair SR components at optical to hard X-ray energies, CR components from primaries at energies up to several GeV and SSC components from pairs at very-high energy (VHE) extending to a TeV for the Crab pulsar and energetic MSPs.  While the SSC component for the Crab peaks in the spectral energy distribution (SED) around 1 - 10 GeV and contributes to the {\it Fermi} spectrum, the SSC components from MSPs such as PSR B1821-24 and B1937+21 have SED peaks at higher energies near 100 GeV.  The MSP SSC components are however suppressed by Klein-Nishina effects and are not predicted to be detectable by current VHE telescopes, but could eventually be detectable by CTA.  The pairs produced by MSPs have much higher energies than those of young pulsars since in their low magnetic fields the pair-producing photons must have higher energies.
Harding et al. (2018) extended the model of Harding \& Kalapotharakos (2015) to both lower (IR and optical) and higher (up to 100 TeV) photon energies to reveal an additional spectral component above 10 TeV from primary particles inverse-Compton scattering pair SR.  Such a VHE component could account for the recent pulsed emission detected by HESS above 7 TeV from the Vela pulsar (Djannati-Atai 2017).  Similar 10 TeV emission is also expected from Crab-like pulsars and energetic MSPs (Harding et al. 2019).   

While global PIC models also show that most particle acceleration is located in the current sheet, the surface magnetic fields and particle energies they can treat numerically are limited to $B \lsim 10^{6}$ G and $\gamma \lsim 10^3$ to resolve the plasma frequency, cyclotron frequency and skin depth.  Since this magnetic field and energy are well below those of pulsars, it is necessary to scale-up the simulation results to model high-energy emission.  There have generally been two approaches to this problem, neither completely satisfactory.  The approach of Cerutti et al. (2016) and Philippov \& Spitkovsky (2018) is to model the synchro-curvature radiation from the PIC-energy particles, which will be in the SR regime given the very low particle energies, and scale-up the photon spectra to GeV energies.  However, it is not clear that the PIC-modeled particle dynamics will be the same as it would be if the particles were to be accelerated to the much higher realistic pulsar energies.  Also, the particles with higher energy and small pitch angles will be in a different regime of synchro-curvature emission and may be closer to the CR regime.    The alternate approach of Kalapotharakos et al. (2018) is to scale up the surface magnetic field and consequent $E_\parallel$ and particle energies to those of a real pulsar on trajectories that are calculated in parallel with those of the PIC particles.  In this case, they find that particles reach  $\gamma \sim 10^7 - 10^8$ in the radiation-reaction limited regime where the acceleration is balanced by CR losses.  The CR spectra reach $\gamma$-ray energies with cutoffs around a few GeV, similar to those of {\it Fermi} pulsar spectra.  The drawback to this approach is the need to artificially suppress the particle pitch angles at low altitudes since their gyro-motion is very small and cannot be resolved by PIC models.  

\subsection{Light curve modeling}
\label{sec:LC}

A variety of the above emission models have been used to fit multi-wavelength light curves of MSPs.  Venter et al. (2009) used geometric versions of outer gap (OG) and slot gap models, as well as the PSPC model to perform by-eye fits to both $\gamma$-ray and radio light curves to eight MSPs with detected $\gamma$-ray pulsations.  They used the two-pole caustic (TPC) model (Dyks \& Rudak 2003) as the geometric version of the slot gap.  The radio light curves were fit with cone beams centered on the magnetic poles at relatively low altitude (Story et al. 2007). It was found that most of the MSP light curves in their sample were best fit with TPC or OG models, while only two were better fit with PSPC models.  The  TPC/OG fits were MSPs whose $\gamma$-ray peaks lagged the radio peaks (Class I), while the MSPs fit by PSPC models had radio peaks that lagged the $\gamma$-ray peaks (Class III), thus identifying two of the three distinct classes of MSP high-energy light curves.  Several years later, {\it Fermi} discovered pulsations from MSPs whose $\gamma$-ray peaks were aligned in phase with their radio peaks (Guillemot et al. 2012) (Class II).  Some of these MSPs were already known to have non-thermal X-ray peaks that were also aligned with their radio peaks.  Venter et al. (2012) (see also Johnson et al. 2014) fit the $\gamma$-ray and radio light curves of several MSPs of this class with altitude-limited OG and TPC models (see Figure \ref{fig:J0030}), where the emission in the gaps was limited between minimum and maximum radii.  The radio emission of these pulsars was found to be at much higher altitude compared to the cone beam emission near the polar caps and formed caustics similar to those of the $\gamma$-ray emission.  The Class II MSPs thus have light curve characteristics close to that of the Crab pulsar and the Crab-like PSR B0540-69 in the Large Magellanic Cloud, whose radio, optical, X-ray and $\gamma$-ray peaks are all aligned in phase.  It is likely that for the short periods and small magnetospheres of both Crab-like pulsars and MSPs, the radio emission occurs at large altitudes relative to the light cylinder where the emission forms caustics that are in phase with the high-energy emission.

\begin{figure}[t]
\includegraphics[width=60mm]{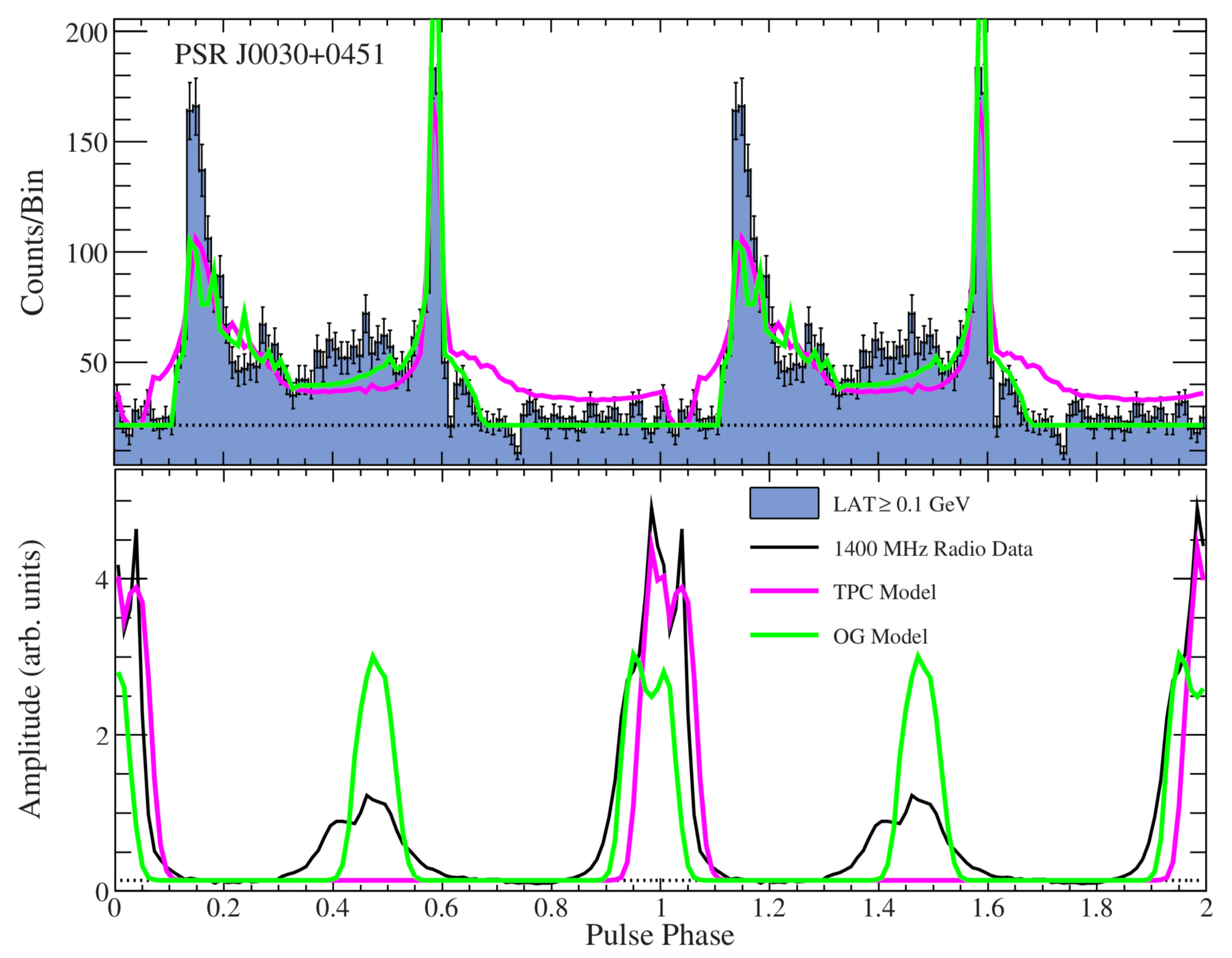}\includegraphics[width=60mm]{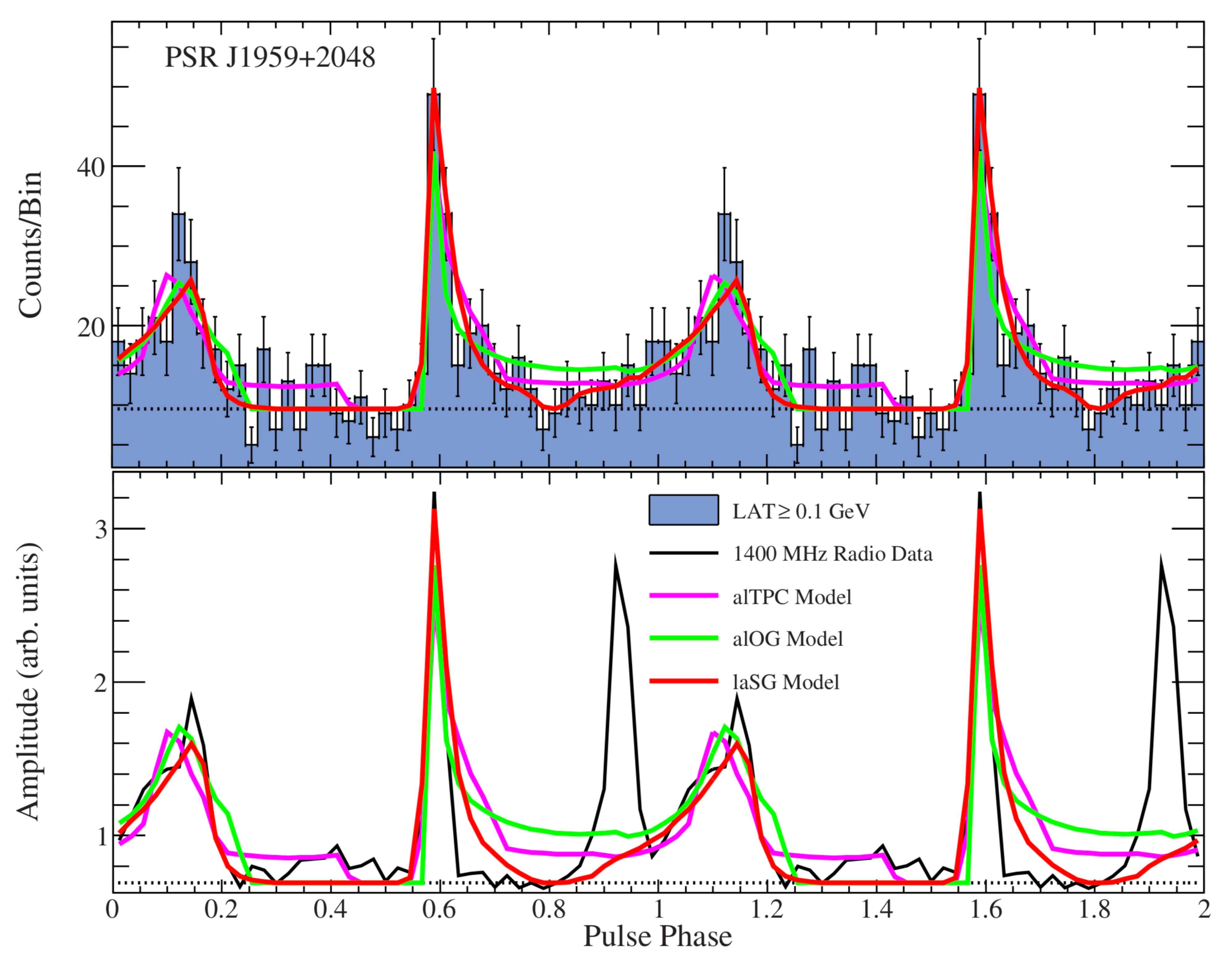}
\caption{Examples of joint $\gamma$-ray (top) and radio (bottom) light curve fits of two {\it Fermi} MSPs.  J0030+0451 (left) is a Class I MSP while J1939+2048 (right) is a Class II MSP having both polar cap and caustic radio peaks. From Johnson et al. (2014).}
\label{fig:J0030}       
\end{figure}

Johnson et al. (2014) performed more formal fits of the same OG, TPC, PSPC and altitude-limited OG and TPC models to 40 MSP light curves of the {\it Fermi} second $\gamma$-ray pulsar catalog (Abdo et al. 2013).  Using a maximum Likelihood technique, they found that Class I and Class II MSP light curves are best fit with either OG or TPC models in about equal numbers while Class III MSPs were exclusively fit with PSPC models.  The MSP light curves that prefer OG models tend to have little or no off-peak emission (See Figure \ref{fig:J0030}, left) while those that prefer TPC models tend to have significant levels of off-peak emission and/or two equal widely spaced peaks.  In this sample, the Class II MSPs with aligned radio and $\gamma$-ray peaks tend to have the shortest periods and the highest surface fields and spin-down power.  The Class III MSPs tend to have the lowest spin-down power.  The light curve fits also showed that MSPs have a wider distribution of inclination $\alpha$ and viewing angles $\zeta$ than normal pulsars whose $\alpha$ and $\zeta$ values are higher (Pierbattista et al., 2015).  Chang et al. (2018) fit the light curves of Class II MSPs using altitude-limited TPC and energy-dependent OG models in VRD magnetic fields but added a perturbation factor to modify the fields due to effects of currents.  

Several of the Class II MSPs, notably PSR B1821-24 (Johnson et al. 2013) and J1939+2048 (See Figure \ref{fig:J0030}, right), not only have radio peaks that are aligned with X-ray and  $\gamma$-ray peaks but also radio peaks that are not aligned with high-energy peaks but are possibly at phases of magnetic poles.  This suggests that MSPs can have both normal radio components above the polar caps as well as high-altitude caustic components.  These two types of radio emission can be identified by their predicted distinctive polarization, with polar cap radio components showing RVM-like S-shaped position angle swings and high polarization degree while the caustic components show fast sweeps of position angle and depolarization at the peaks (Dyks et al. 2004).  This behavior is in fact seen in the radio polarization of PSR B1821-24 (Bilous et al. 2015), where the high-energy aligned radio peaks show caustic polarization and the polar cap radio peak shows RVM polarization.  These MSPs are similar to the Crab pulsar which has both caustic radio peaks aligned with high-energy peaks and a polar cap radio precursor.  

Although OG and SG/TPC models each have some of the elements that can match observed $\gamma$-ray pulsar characteristics, neither model is able to satisfactorily fit all the MSP light curves (Venter \& Harding 2014).  This result was also found for fits of OG and TPC models to the light curves of young {\it Fermi} pulsars (Pierbattista et al., 2015).  Ultimately, the reason for the failure of OG and SG/TPC model light curve fitting is that neither is the correct physical model for pulsar high-energy emission.  As discussed above, global dissipative MHD and PIC simulations showed that the main emission is not inside the light cylinder but from the current sheet.  The original OG and TPC models were not able to include the current sheet emission since they used the VRD whose magnetic field structure requires particle trajectories along field lines outside the light cylinder to be super-luminal.  On the other hand, in force-free and dissipative magnetospheres particle trajectories along magnetic field lines remain sub-luminal at all radii.  Furthermore, the model light curves of emission using MHD and PIC global magnetospheres combine some of the best elements of both OG and SG/TPC models, such as very low off-peak emission, bridge emission between the peaks and emission from both hemispheres.

\begin{figure}[t]
\includegraphics[width=120mm]{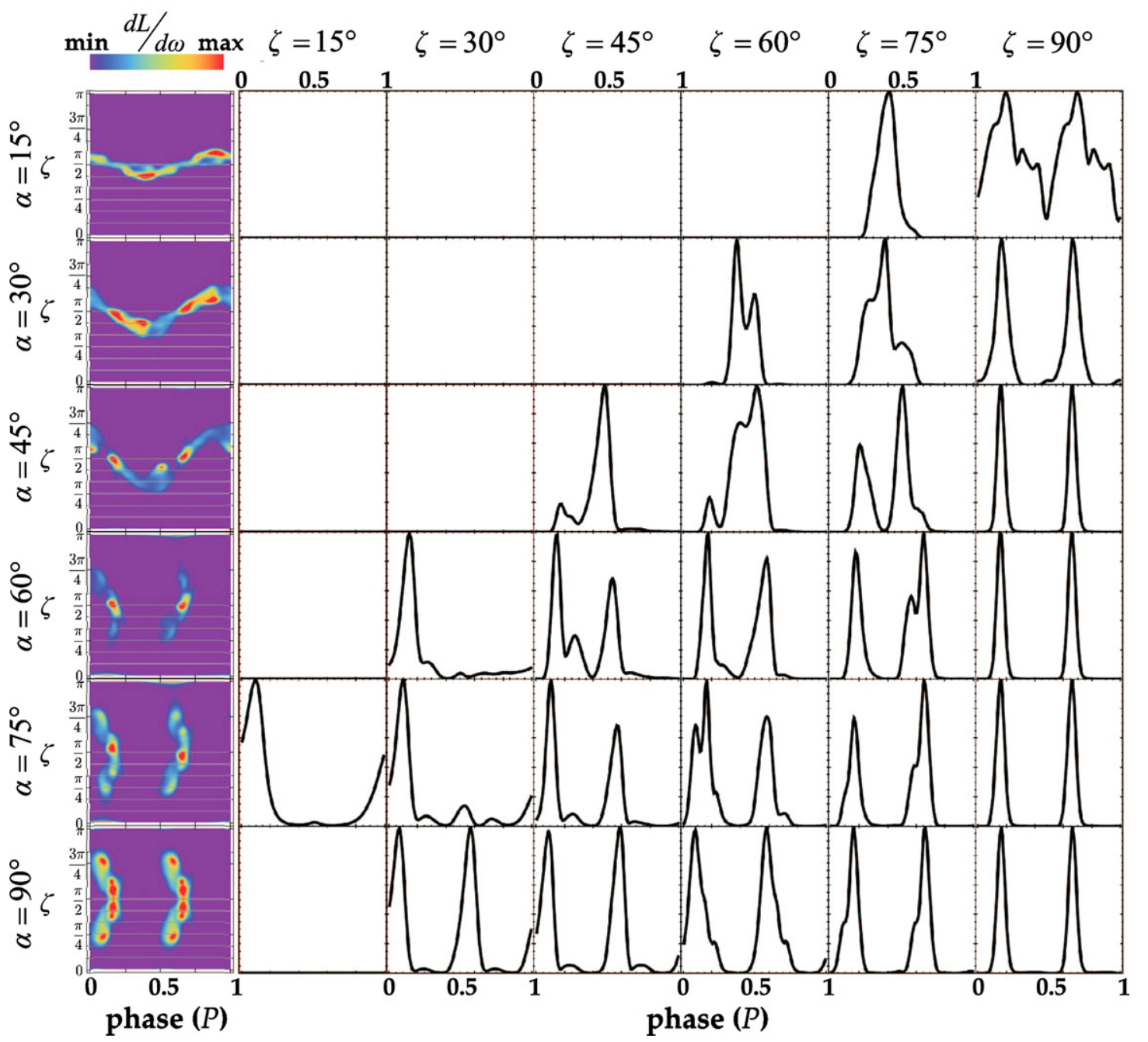}
\caption{Skymaps of high-energy luminosity per solid angle $dL/d\omega$ at different inclination angles $\alpha$ and light curves at observer angles $\zeta$, for a dissipative MHD magnetosphere model, FIDO, having infinite conductivity (force-free) inside the light cylinder and a finite conductivity outside. From Kalapotharakos et al. (2014).}
\label{fig:FIDO}       
\end{figure}

Bai \& Spitkovsky (2010) modeled the geometry of high-energy emission and light curves using FF magnetosphere models assuming emission along the last open field line, or separatrix, that extended from the neutron star surface to beyond the light cylinder into the current sheet.  The caustics that appear in the emission sky maps in this "separatrix model", as in other current sheet emission models, are stagnation caustics formed as the particles near and outside the light cylinder must drift backward along field lines to counteract the field sweepback. The particles are thus able to maintain sub-luminal velocities with their trajectories parallel to field lines as they move out nearly radially.  The emission from these particles at different radii in the current sheet will all arrive in phase to form the caustics.  Depending on magnetic inclination and viewing angle, light curves can have one or two peaks of varying separation.  In the double-peaked light curves, the peaks come from opposite hemispheres similar to those of TPC models and unlike the one-hemisphere OG models.

Sky maps (distribution of emission as a function of observer angle and phase with respect to the rotation axis) and light curves from dissipative magnetospheres were explored by Kalapotharakos et al. (2014) who used the self-consistent $E_\parallel$ distribution of the models to accelerate particles and emit CR emission.  It was found that models with infinite $\sigma$ inside $R_{\rm LC}$ and large but finite $\sigma$ outside $R_{\rm LC}$ or FIDO (Force-free Inside Dissipative Outside) models were able to match the light curve characteristics of young {\it Fermi} pulsars, particularly the $\gamma$-ray peak separation vs. radio lag.  Importantly, matching the radio lags requires the distribution of emission in the 3D current sheet to be non-uniform and non-axisymmetric.  Models with uniform emission throughout the current sheet produce larger radio lags not compatible with the data.
Cao et al. (2019) also modeled light curves from dissipative magnetosphere models calculated using a pseudo-spectral method (Cao et al. 2016), different from the MHD methods used by Kalapotharakos et al. (2012a) and Li et al. (2012).  They explored both geometric models with uniform emissivity along fields and uniform $\sigma$, and emission from particles accelerated by the self-consistent $E_\parallel$ from a FIDO emissivity distribution.  A succession of sky maps and light curves for dissipative magnetospheres with uniform increasing $\sigma$ show that light curve peaks progressively move to later phase with $\sigma$ increasing from zero (vacuum) to near-FF, confirming the previous results of Kalapotharakos et al. (2012b).  Since uniform emissivity and $\sigma$ models do not match {\it Fermi} radio lags, they confirm that FIDO models with non-uniform emission in the current sheet is a better match to the data.

\begin{figure}[t]
\hspace{-0.4cm}\includegraphics[width=120mm]{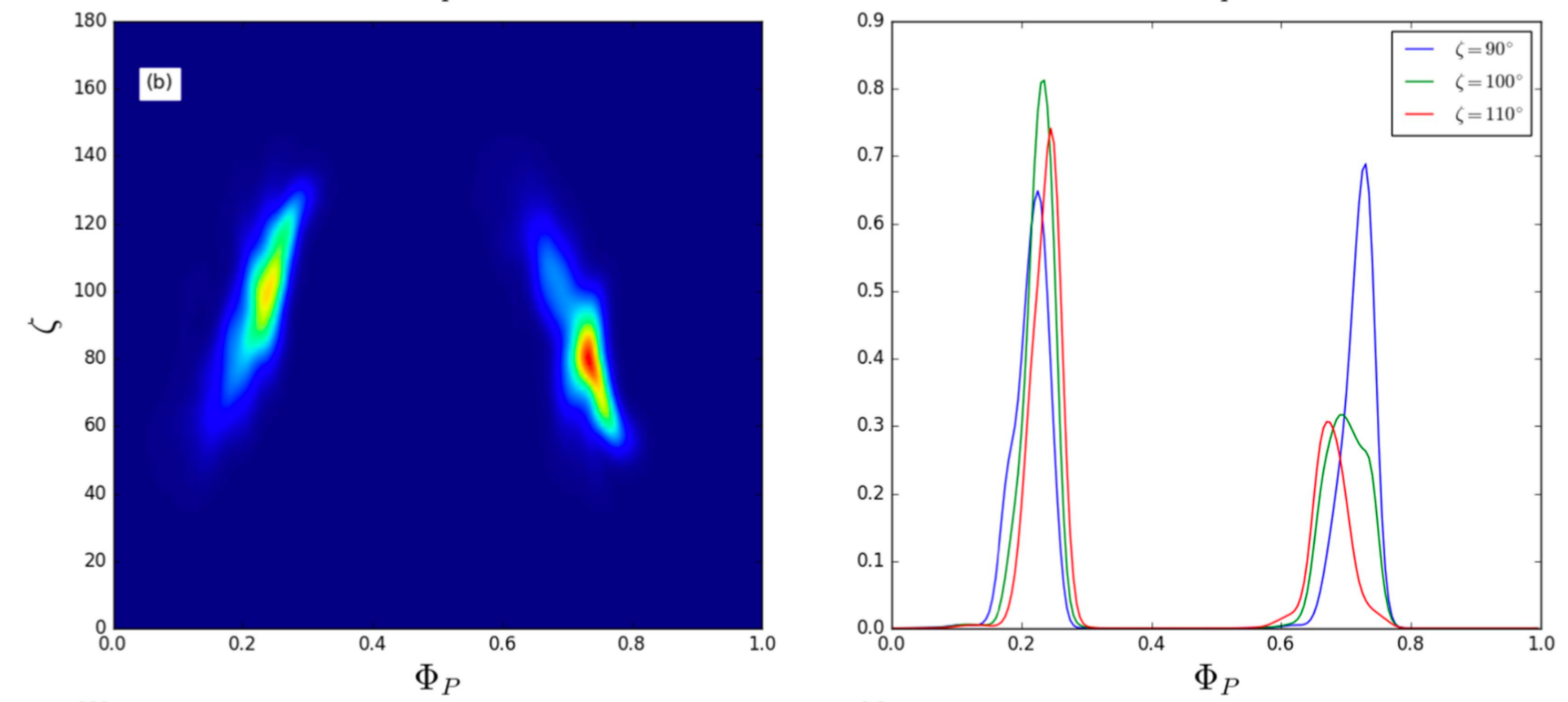}
\hspace{0.2cm}\includegraphics[width=56mm]{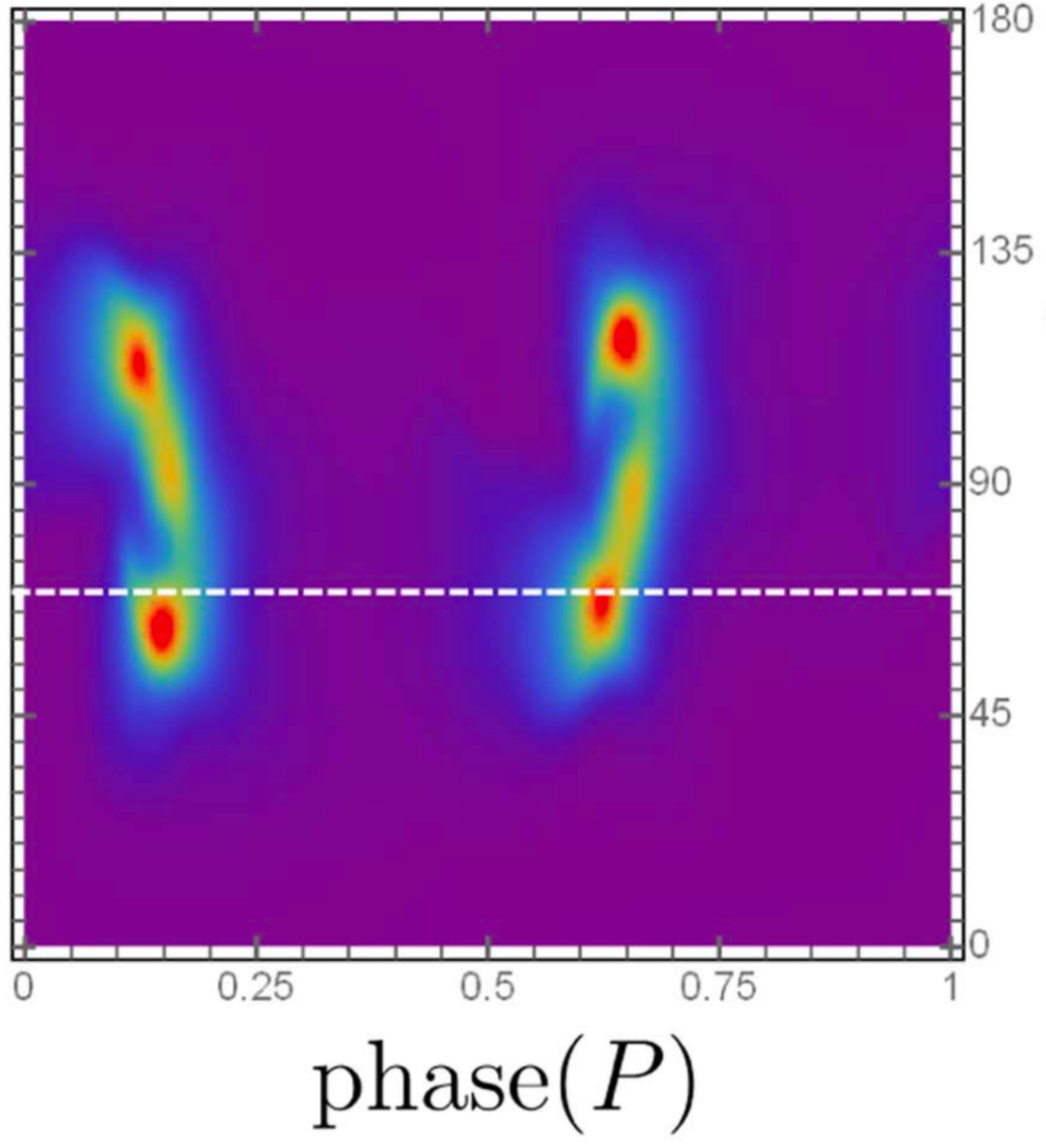} \hspace{0.4cm}\includegraphics[width=56mm]{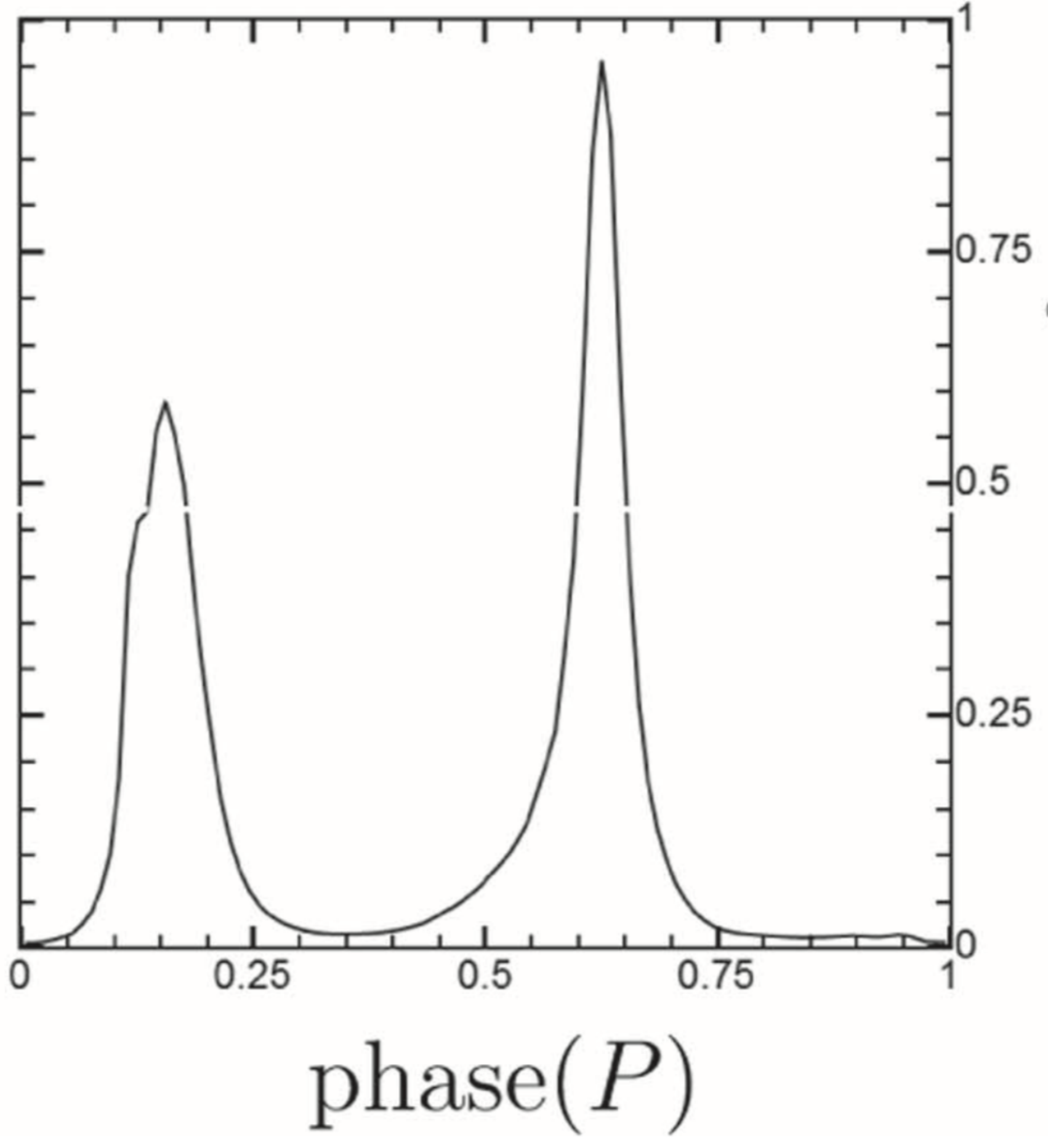}
\caption{High-energy sky maps (left) and light curves (right) from PIC current sheet models.  Top row is for $\alpha = 60^\circ$ and light curves are for three values of $\zeta$ from Philippov et al. (2018).  Bottom row is for $\alpha = 75^\circ$, and light curves are for $\zeta = 70^\circ$ from Kalapotharakos et al. (2018).  Phase 0 and 0.5 in these plots are the phases of the magnetic poles.}
\label{fig:PIClc}       
\end{figure}

The FIDO dissipative models do a good job matching the radio lags of young pulsar light curves assuming that the radio phase is equal to the magnetic pole phase, which would be true if the radio emission occurs relatively near the neutron star surface.  While this assumption may be justified for young pulsars, the {\it Fermi} MSPs have larger radio lags on average that do not fit FIDO models as well.  This is likely because in the smaller magnetospheres of MSPs the radio emission height is a large fraction of $R_{\rm LC}$, with the phase shift from aberration and retardation adding to the $\gamma$-ray model phase lag from the magnetic pole.  If the $\gamma$-ray model light curves were known to be accurate, the observed MSP radio lags would give us information on their radio emission heights.  This may be possible as the $\gamma$-ray models improve.  Additionally, the Neutron Star Interior Composition Explorer (NICER) is accurately measuring phases of thermal X-ray peaks of MSPs, which locate the phase of the magnetic poles, relative to the radio peak phases (Ray et al. 2018).  Using NICER observations in conjunction with {\it Fermi} and radio data will further constrain both $\gamma$-ray and radio models.

The sky maps and high-energy light curves  in PIC models are similar to those of the striped wind and current sheet models, showing one or two peaks with varying phase separation, but are sensitive to the assumptions made in artificially scaling up the PIC particle energies to those of real pulsars, as discussed in Section \ref{sec:CS}.  The $\gamma$-ray light curves computed from scaling up SR of PIC particles to $\gamma$-ray energies (Cerutti et al. 2016, Philippov \& Spitkovsky 2018), shown in Figure \ref{fig:PIClc}, have model phase lags from the magnetic pole that are larger than the radio lags of young pulsars, but might better agree with those of MSPs.  That would imply however, that MSP radio emission occurs at the neutron star surface.  The light curves computed from CR of  PIC particle with scaled-up energies (Kalapotharakos et al. 2018), also shown in Figure \ref{fig:PIClc}, have smaller model phase lags from the magnetic pole that agree better with radio phase lags of young pulsars and even allow for some additional lag from aberration and retardation of radio emission located above the neutron start surface.  Interestingly, the sky maps and light curves of the PIC models of Kalapotharakos et al. (2018) agree with those of the MHD FIDO models that do not need to be scaled up since they can be computed for realistic pulsar fields.  This implies that CR emission models are consistent for both dissipative MHD and PIC approaches, with the caveat that the dissipative models must assume an arbitrary $\sigma$ distribution.

\section{Pairs and Death Lines}
\label{sec:pairs}

The production of electron-positron pairs is believed to be an essential element of pulsar physics, providing the plasma to fill the magnetosphere and supply the wind particles, the charges and the currents, as well as the pairs needed for coherent radio emission.  There are several sites in the magnetosphere where pair production and cascades can occur: at the polar caps, in the outer gaps, in the current sheet and at the Y-point, where the separatrices from opposite poles merge.  Since the latter three sites can only supply pairs on a limited range of field lines near the separatrix, polar cap pair cascades are required at minimum to supply pairs throughout the rest of the magnetosphere and particularly to facilitate the radio emission above the polar caps.  Electromagnetic pair cascades at the polar caps (Sturrock 1971, Daugherty \& Harding 1982) are enabled through the QED process of one-photon pair production (Erber 1966), where a single photon can convert to an electron-positron pair in a magnetic field that is strong enough to supply the additional momentum required.  The pair production threshold, $\epsilon > 2mc^2 / \sin\theta$, where $\epsilon$ is the photon energy in $mc^2$, requires the photon momentum to have an angle $\theta$ to the local magnetic field.  Since relativistic particles accelerated along the field lines radiate photons within a narrow angle of their direction, the photons must travel some distance across curved field to acquire the angle to create a pair.  The particle acceleration length plus the physics of pair production thus sets the length scale of the polar gaps since once pairs are created they screen the surface $E_\parallel$ and limit the particle acceleration.  The polar cap pair cascades can be initiated by either CR or by inverse Compton photons from accelerated particles scattering thermal X-ray from the neutron star surface (Zhang et al. 1997b, Harding \& Muslimov 1998, 2002, Zhang \& Harding 2000).  The pairs will be produced with non-zero pitch angles and will radiate SR that can produce more generations of pairs until the photons eventually can escape the magnetosphere.  The total number of cascade pairs per primary particle is the pair multiplicity which can reach up to $\sim 10^4$ in dipole fields.

Outer gap pair cascades must rely on photon-photon pair creation since the magnetic fields in the outer magnetosphere are too low for one-photon pair creation.  The CR by primary particles accelerated by the gap interact with thermal X-rays from the neutron star surface, or non-thermal X-rays from the gap cascade in the case of Crab-like pulsars, to produce secondary pairs that then radiate SR.  The pairs screen the gap $E_\parallel$ to limit the width and voltage of the gap.  In the original outer gap model (Cheng et al. 1986), the current through the gap is sustained entirely by the gap pair cascades so that the outer gap dies if the gap cascades cannot take place.  However, Hirotani \& Shibata (2001) noted that the self-sustained gap current alone cannot account for the observed luminosity of $\gamma$-ray pulsars and proposed that there must be external currents in the outer gaps.  

As pulsars age and spin down, the gap voltage drops below that required to radiate pair-producing photons.  One can define boundaries in period $P$ and period derivative $\dot P$ space that define a "death line" for pair production in the gaps.  The location of the death line will depend on the particular gap model, with the outer gap death lines in $\dot P$ vs. $P$ space (Zhang et al. 2004) located above those of polar cap death lines.  Since pairs are thought to be a requirement for radio emission, the polar cap pair death line has been adopted as the death line for pulsar radio emission.  Given the angle dependence of the one-photon pair threshold, the curvature of the near surface magnetic field is important in determining the  polar cap death line.  In a pure dipole field, the death line for producing pairs by CR occurs at a pulsar age $\sim 10^7$ yr for normal pulsars which is much less than the maximum age $\sim 10^8$ yr of observed radio pulsars (Ruderman \ Sutherland 1975, Harding \& Muslimov 2002).  In the MSP population, the CR pair death line is even more restrictive with only a handful of MSPs lying above the line.  Although Zhang, Harding \& Muslimov (2000) found that the death line for producing pairs by IC photons lies below the entire normal and MSP populations, IC pair cascades produce very low pair multiplicity (Harding \& Muslimov 2002) that may not be sufficient for radio emission.  It has therefore been suggested that multipole fields are present near the neutron star to enable CR pair cascades in older pulsars and MSPs (Arons 1986).  Multipolar fields have a smaller field line radius of curvature and can produce larger $E_\parallel$ than dipole fields, both of which allow more robust pair cascades (Harding \& Muslimov 2011).  

More recent global models of the pulsar magnetosphere have substantially changed the earlier models assuming steady-state polar cap and outer gap cascades.  The FF solutions showed that the global magnetosphere requires a current distribution along open magnetic field lines that is different from the Goldreich-Julian current density $J_{\rm GJ} = \rho_{\rm GJ} c$ assumed by previous polar cap and outer gap models (Timokhin 2006).  In the global FF magnetosphere, regions of current density $J > J_{\rm GJ}$ (super-GJ),  $0 < J < J_{\rm GJ}$ (sub-GJ) and $J < 0$ (anti-GJ) exist across the polar caps, with a distribution that depends on magnetic inclination.  In this picture, the main current ($J > 0$) flows out over most of the polar cap while the return current ($J < 0$) occurs in regions that connect to the (spin) equatorial current sheet.  In order for polar cap accelerators and the pair cascades they produce to provide the currents required by the global model, they must be time-dependent rather than steady-state.  Time-dependent models of vacuum gap pair cascades were developed by Timokhin (2010) and for Space-Charge Limited Flow (SCLF) gaps by Timokhin \& Arons (2013), showing that in both cases the accelerators produce bursts of pairs that fully screen the electric field, halting acceleration until the pair cloud exits the gap, followed by renewed acceleration and another burst of pair production.  In non-steady polar gaps the $E_\parallel$ is higher than in steady gaps, so that the maximum pair multiplicity for CR cascades can reach a higher value $\sim 10^5$ (Timokhin \& Harding 2015).  

Some PIC models of pulsar magnetospheres have attempted to include pair production where PIC particles reach a specified threshold energy (Chen \& Belodorodov 2014, Philippov et al. 2015).  Although such an approach cannot resolve the actual microphysics of pair production, it shows the possible locations in the global magnetosphere where pair production could take place.  These locations are places where particles are accelerated to high energies such as above the polar caps, at the Y-point, and in the current sheet.  The simulations suggest that the polar caps can supply pairs and/or currents along field lines with super-GJ current $J > J_{\rm GJ}$, but in the return current regions with anti-GJ current $J < 0$, pairs from the Y-point and current sheet may be needed.

\section{Thermal X-ray Emission and Field Structure}
\label{sec:PC}

The surface thermal emission from MSPs, which are well past the age when neutron star cooling would be important, is most likely caused by heating by back-flowing energetic particles from pair cascades at the polar caps or in the outer magnetosphere, and from global return currents .  Early models of pulsar polar cap acceleration concluded that some flux of particles returning to the neutron star surface was inevitable, since the electrons and positrons of the pairs that are produced by the upward accelerating particles will move in opposite directions in the electric field.  Since the downward-moving particles, produced near the top of the gap, will gain nearly the full gap voltage before they reach the surface, this energy will be deposited in the NS surface layers and heat the polar cap.  Vacuum gaps (e.g. Ruderman \& Sutherland 1975) that spontaneously break down through pair cascades produce equal numbers of energetic upward and downward particles, and thus a higher level of polar cap heating than the SCLF gaps (e.g. Arons \& Scharlemann 1979) that return only a small fraction of positrons to screen the electric field.  The steady SCLF gaps were shown to produce polar cap heating temperatures and luminosities that are in agreement with observed thermal X-ray components (Arons 1981, Harding \& Muslimov 2001, 2002), whereas the vacuum gaps can produce polar caps much hotter than observed.  
Outer gaps (Cheng et al. 1986) break down in a way similar to that of polar vacuum gaps, returning a flux of energetic particles to the NS that is equal to the flux of upward energetic particles.  Halpern \& Ruderman (1993) noted that a flux of returning particles from outer gaps that was equal to the flux of outward-going particles needed to produce the $\gamma$-ray emission of the Geminga pulsar would heat the polar caps to a temperature much higher than observed, suggesting that some mechanism for reprocessing the thermal emission from heating is needed (see also Wang et al. 1998).  
Even in the non-steady SCLF case, there are nearly equal fluxes of upward and downward moving particles, and thus high levels of polar cap heating, during the pair formation and screening part of the cycle (Timokhin \& Arons 2013).  The average polar cap heating luminosity then depends on the duty cycle of the time-dependent pair cascades, which depends on the time required for the pair cloud from each cycle to clear the accelerator.  Although the pair cascade duty cycle and thus the actual polar cap heating luminosity is difficult to assess with present calculations, it is possible to estimate by assuming that the pair cloud exits the gap on a timescale $\sim R/c$, where $R$ is the neutron star radius.  This approximate heating gives surface temperatures that agree with those observed for both normal and MSPs (Timokhin \& Harding 2019).  

The pattern of polar cap heating will depend on the current distribution across the polar cap which, as discussed above, is a function of magnetic inclination.  It also depends on what type of accelerator (vacuum or SCLF) is operating.  With free supply of both signs of charge, thought to be the case for all except magnetar-strength magnetic fields (Medin \& Lai 2007), SCLF gaps will operate.  For time-dependent SCLF gaps (Timokhin \& Arons 2013), pair cascades (and polar cap heating) occur only in regions with super-GJ or anti-GJ current.  No pair cascades (and no polar cap heating) occur in regions of sub-GJ current since a flow of low-energy primary electrons alone can supply the current.  For aligned rotators, most of the polar cap has sub-GJ current and is not heated, surrounded by a ring of anti-GJ return current adjacent to the last open field lines.  As the inclination increases, the region of anti-GJ current increases toward the direction of the spin equator (and current sheet) while the region of sub-GJ current shrinks.  At inclinations above around $45^\circ$, a region of super-GJ current appears and grows on the equatorial side of the anti-GJ current, while the sub-GJ region continues to shrink.  At inclination of $90^\circ$, a region of super-GJ current covers nearly the entire polar cap.  Thus, for most inclination angles, the heated area is either nearly symmetric around the outside of the polar cap or covers a large part of the polar cap that is nearly symmetric about the poloidal plane containing the magnetic and spin axes.  

The heated areas inferred from observed thermal components of MSPs are typically much smaller than the polar cap area (Becker 2009).  A study of the 2D heated area of centered and offset  polar caps for the case of a steady-state SCLF gap (Harding \& Muslimov 2011) shows that the hot spot radius is predicted to be much smaller than the canonical polar cap radius $R_{\rm PC} = R (\Omega R/c)^{1/2}$.  The hot spot radius is $R_{\rm hot} \sim 0.03 - 0.1\,R_{\rm PC}$, which for MSPs is $R_{\rm hot} \sim 50 - 100$ m.  Most of the heating occurs near the center of the polar cap because the gap voltage is largest near the magnetic axis, where the radius of curvature is largest and screening by cascades from CR photons is least effective.  Observed thermal components of MSPs typically have hot spot radii of around 100 m (Bogdanov et al. 2008, 2013), in agreement with the theoretical heated area.  Thermal X-ray light curve fitting of several MSPs indicates that the hot spots are offset from the magnetic axis (Bogdanov et al. 2007), implying offset dipoles or multipoles.

\begin{figure}[t]
\hskip -0.3cm
\includegraphics[width=60mm]{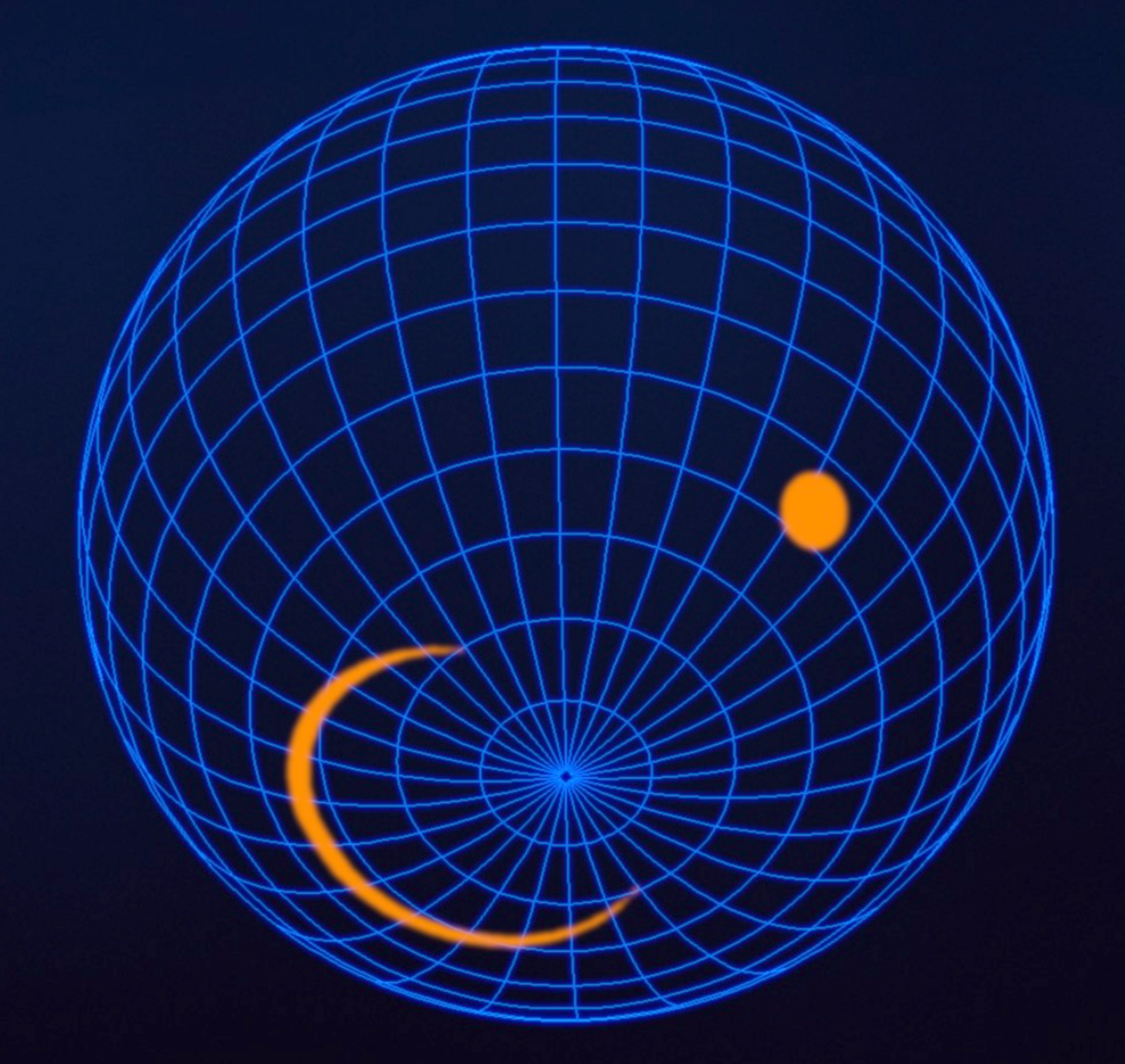}\includegraphics[width=58.5mm]{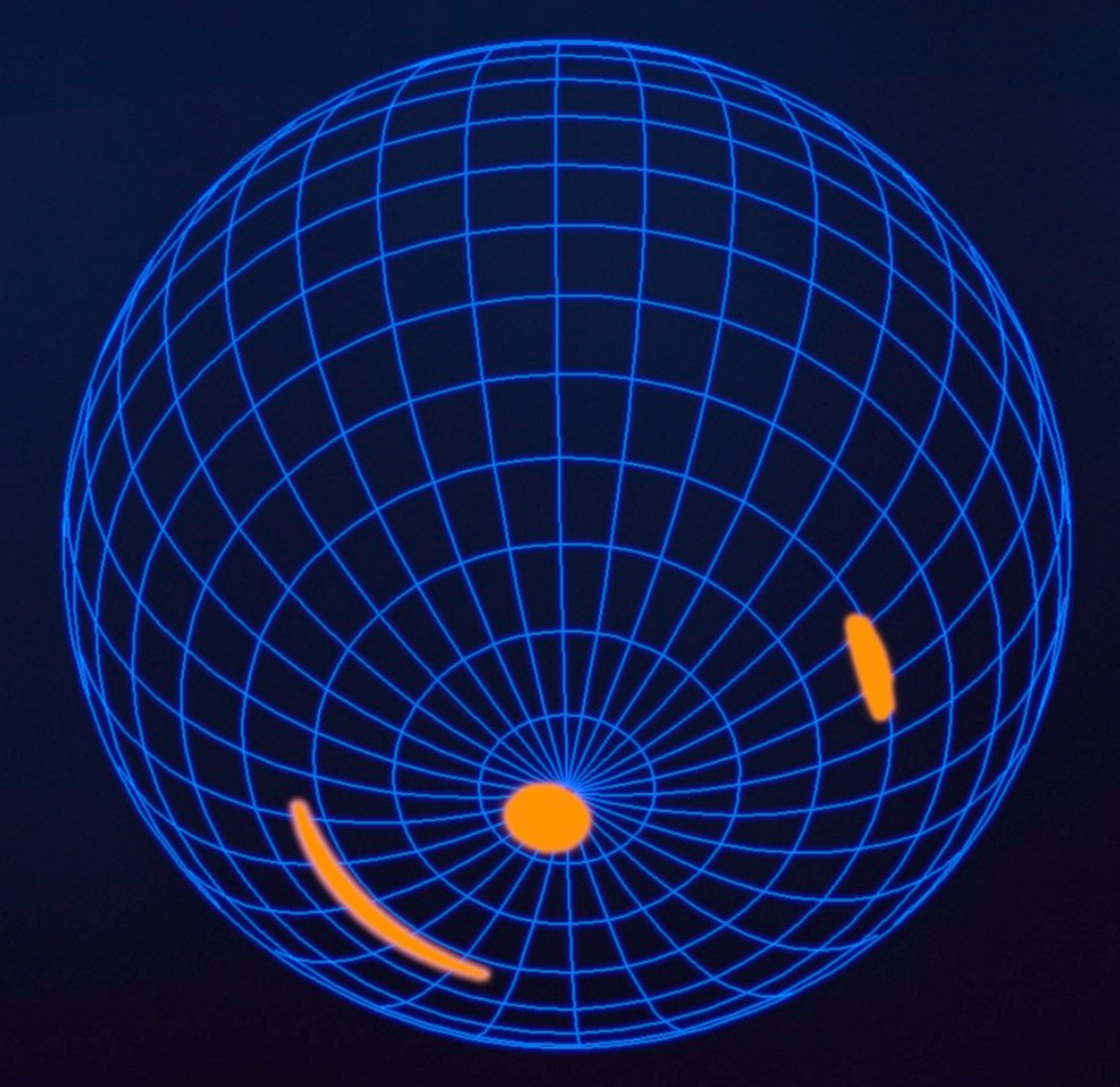}
\caption{Heated spots viewed $144^\circ$ from the rotation axis, as inferred from fitting of the NICER X-ray waveforms of PSR J0030+0451 by two independent groups: Riley et al. (2019) (left panel), Miller et al. (2019) (right panel).  Illustration credit: NASA Goddard Space Flight Center.}
\label{fig:J0030}       
\end{figure}

Recent fitting of the thermal X-ray light curves of MSP J0030-0451 observed by NICER to obtain constraints on the neutron star mass and radius (see Chapter 11) require two or three heated spots all in the same hemisphere (relative to the spin axis), with the observer direction in the opposite hemisphere (Miller et al. 2019, Riley et al. 2019).  The spots are thus not antipodal as would be expected for a star-centered dipole field.  Furthermore, one spot is much smaller than the polar cap size and the other main spot is a very extended crescent or ellipse.  These results strongly suggest that this MSP has a non-dipolar and very offset magnetic field (Bilous et al. 2019).  Gralla et al. (2017) showed that a combined dipole plus quadrupole field can produce hot spots similar to what is inferred from the NICER data, but the spots are antipodal since the fields were star centered.  Chen et al. (2020) explored a vacuum NS-centered dipole plus a quadrupole offset along the spin axis by $0.4 R$ whose polar caps resemble the hot spot shapes found by Riley et al. (2019).  They then used a global FF  model, matching model $\gamma$-ray light curves to Fermi data to determine the magnetic inclination angle which is not constrained by the NICER waveform modeling alone.  Kalapotharakos et al. (2020) took a different approach, first determining configurations of dipole and quadrupole components, both with arbitrary offset and orientation, whose hot spots produce thermal X-ray light curves that accurately fit those of  NICER.  They found several degenerate field configurations with different offsets, orientations (including dipole inclination angle) and quadrupole-to-dipole ratio that all match the NICER data.  Using these solutions, they then computed the $\gamma$-ray light curves in a FF magnetosphere to compare with those of Fermi, finding a configuration with magnetic inclination angle $\sim 90^\circ$, dipole offset $\sim 0.4 R$ and quadrupole offset $\sim 0.3 R$ that is compatible with both NICER and Fermi data.  In this field configuration,  the surface quadrupole $B_{\rm Q}$ at the polar caps has a strength that is about 10 times that of the dipole field, which for J0030-0451 is $B_{\rm D} = 2.2 \times 10^8$ G.

\section{Millisecond Pulsars in Binary Systems}
\label{sec:bin}

The majority ($\sim$ 80\%) of rotation-powered MSPs are in binary systems, which is not unexpected given that their progenitors are Low Mass X-Ray Binaries (LMXBs).  A class of binary MSPs are in very close systems in which the pulsar wind is heating the companion star and interacting with its induced wind or magnetosphere to form an intra-binary shock.  This class is divided into two main groups called Black Widows, whose companions are extremely light (a few Jupiter masses) and severely ablated, and Redbacks with heavier (several tenths of Solar masses) but also heated companions.  The number of Black Widow and  Redback MSPs has recently increased from four to nearly 30 with pointed radio searches of {\it Fermi} unidentified $\gamma$-ray sources that discover new MSPs (Ray et al. 2012).  The radio pulsar ephemerides and orbital solution then allow a detection of the $\gamma$-ray pulsations.  In a few cases, optical observations of {\it Fermi} sources have revealed the binary orbital periods by detecting the intensity variations of heated and unheated sides of the companions in these tidally-locked systems (Romani 2012, Romani \& Shaw 2011).  This then allowed successful blind searches for the $\gamma$-ray pulsations (Pletsch et al. 2013, Kong et al. 2012).

\begin{figure}[t]
\hskip -0.3cm
\includegraphics[width=125mm]{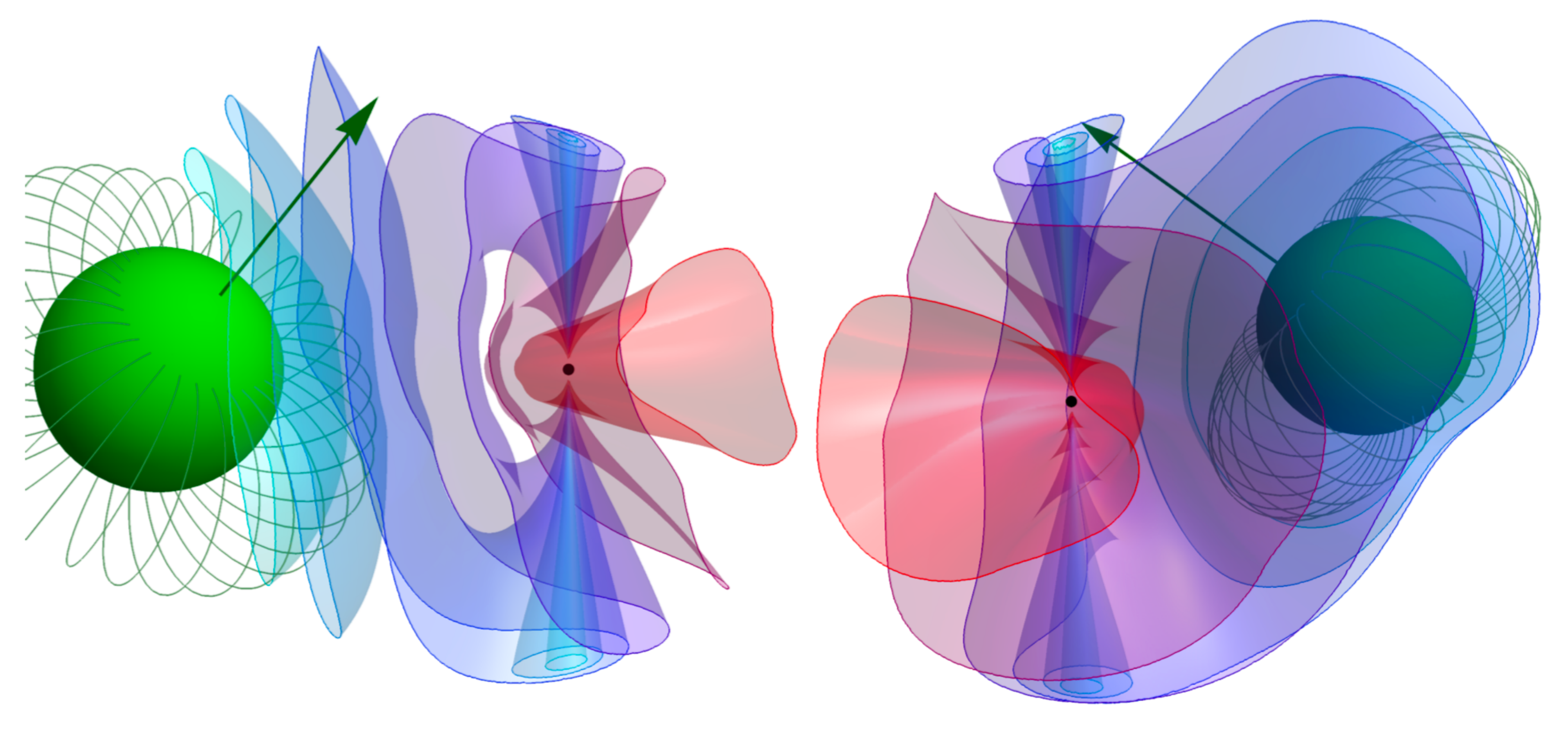}
\caption{Views of isobaric surfaces for an anisotropic wind of a pulsar with magnetic inclination $\alpha = 15^\circ$ and misaligned dipole magnetosphere of the companion.  Isobar colors scale from green to red with increasing surface field strength of the companion star. From Wadiasingh et al. (2018).}
\label{fig:isobar}       
\end{figure}

Many of the Black Widows and Redbacks show non-thermal X-ray emission modulated at the orbital period and two show orbitally-modulated $\gamma$ rays, strongly suggesting that the intra-binary shocks are accelerating particles to ultra-relativistic energies.  Most also show eclipses of the MSP pulsed radio emission that indicate absorption or scattering by the companion wind or atmosphere.  The intra-binary bow shock forms along the pressure balance surface between the pulsar wind and companion star wind or magnetosphere (Harding \& Gaisser 1990,  Wadiasingh et al. 2018) and wraps around one of the two stars depending on which pressure is dominant (see Figure \ref{fig:isobar}).  Both single- and double-peaked X-ray light curves are seen, with the majority of sources showing the X-ray peak(s) at inferior conjunction when the pulsar is in front of the companion, whereas the radio eclipses occur at superior conjunction.  If the X-ray peaks are produced by the radiation of accelerated particles flowing along the shock, then peaks at inferior conjunction require that the shock wraps around the pulsar.  Mildly Doppler-boosted SR from shock-accelerated particles can explain the shapes of the observed X-ray light curves (Wadiasingh et al. 2015, 2017, Romani \& Sanchez 2017).  

\begin{figure}[t]
\includegraphics[width=120mm]{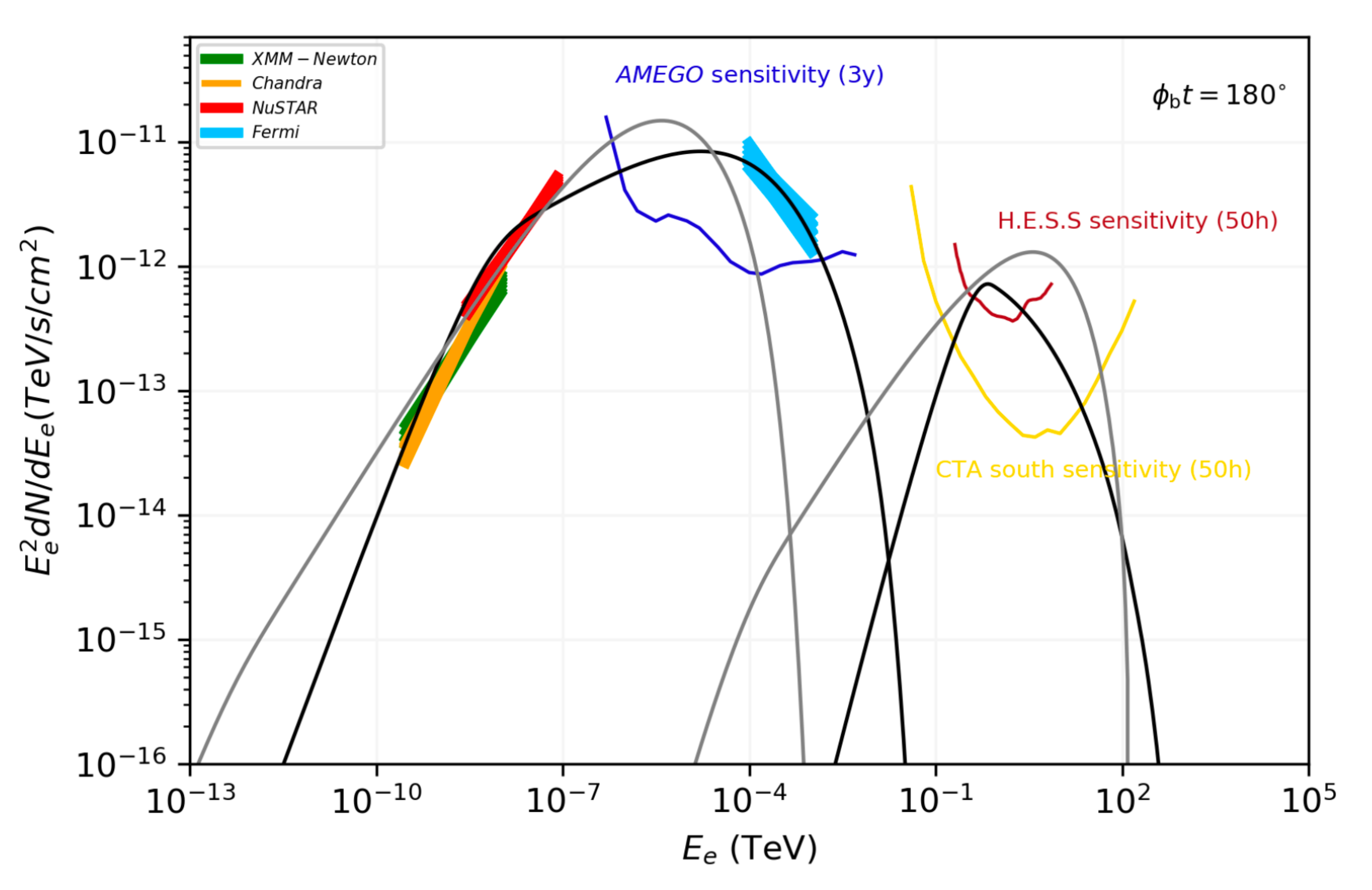}
\caption{Model SR and IC emission components from particles accelerated at the intra-binary shock in the Redback system containing MSP J1723-2837.  Spectra are shown for two models that match the X-ray data but predict different GeV and TeV spectra. From Van der Merwe et al. (2020).}
\label{fig:J1723}       
\end{figure}

MSP binaries are particularly good sources in which to study properties of pulsar wind shocks since they occur much closer to the pulsars than those in pulsar wind nebulae.  The shock-accelerated particle energies could reach as high as several TeV (Harding \& Gaisser 1990) if diffusive acceleration limited by SR losses is occurring.  Reconnection alone will produce lower maximum particle energies that are limited to $\gamma_{\rm max} \sim \gamma_w\sigma_w$, where $\gamma_w$ is the wind bulk flow Lorentz factor and $\sigma_w$ is the ratio of the electromagnetic to particle energy density in the wind (Kandel \& Romani 2019).  If diffusive shock acceleration (DSA) dominates, the particle spectrum will have an index $-\delta$ where $\delta = (r+2)/(r-1)$ and $r$ is the shock compression ratio, if the injected (upstream) spectrum is steeper than $-\delta$.  If the upstream spectrum is flatter than $-\delta$, then the accelerated spectrum retains its slope, with the particles just moving up in energy (Jones \& Ellison 1991).  Since the injected particle spectrum, the pair spectrum from the MSP, has an index $p \sim 1.5$ (Harding \& Muslimov 2011) which is flatter than $\delta = 2$ of a shock with maximum $r = 4$, this will also be the accelerated particle spectral slope.  The expected SR photon spectral index would then be $-(p + 1)/2$ or $\sim 1.25$.  If magnetic reconnection dominates the particle acceleration, the particle spectral index could be as hard as $p \sim 1$ producing a SR spectrum with photon index $\sim 1$.  Since observed power-law spectra of intra-binary shock emission in MSP binaries is $\sim 1$ (Roberts et al. 2015), either of these processes would be compatible.  However, the maximum particle energy could discriminate between DSA and reconnection.  Detection of VHE emission above 1 TeV from IC scattering of shock-accelerated particles on thermal radiation from the companion would argue in favor of DSA or at least some combination of reconnection and DSA (Sironi \& Spitkovsky 2011).  The IC component is predicted to be detectable by Air-Cherenkov telescopes for several MSP binaries (Van der Merwe et al. 2020), as shown in Figure \ref{fig:J1723}.  Both the maximum particle energy and pair multiplicity have implications for a contribution of MSP binaries (Venter et al. 2015) to the energetic cosmic-ray positron excess observed by AMS2 (Accardo et al. 2014).

\section{Outstanding Problems}
\label{sec:probs}

In the area of emission physics, MSPs provide many interesting puzzles that suggest a number of topics for future investigation.  One of the most fundamental questions is: How do MSPs produce electron-positron pairs?  If their magnetic fields are purely dipolar, all but a few known MSPs should not be producing any pairs, much less the high multiplicities required for radio emission and for the $\gamma$-ray pulses from near-FF magnetospheres.  As was discussed in Section \ref{sec:PC}, one proposed solution is that MSPs have more complex multipole fields near their surface.  Higher multipoles can have a local field strength that is much higher than what is inferred from the spin-down rate and will have smaller radii of curvature, both of which enhance magnetic pair creation.  Multipole fields can also produce offset and/or larger polar caps that will increase the surface $E_\parallel$.  There is independent evidence that MSPs have non-dipolar fields from the thermal X-ray profile fits which suggest non-antipodal surface hot spots in several sources.  However, if MSPs really have the magnetic field configuration with the extreme offset inferred by the NICER waveform modeling, what effect does it have on the $\gamma$-ray light curves that are well reproduced by dipole magnetospheres?  There is evidence that the offset field configurations inferred by the NICER data have some effect on the $\gamma$-ray light curves but that the $\gamma$-ray data may also help in narrowing down degenerate  field configurations. What is the geometry and visibility of the radio beams in this case?  Finally, is the inferred field configuration stable?  Another proposed solution is suggested by PIC models of "weak pulsars", those that cannot produce enough pairs to fill their magnetospheres (Chen \& Belodorodov 2014).  Chen at el. (2019) find that pulsars can have global states with intermittent pair production and current flow by which they can reach a near-FF magnetosphere.  Global magnetosphere simulations with non-dipolar fields will be needed to answer some of these questions.

Another unresolved question, that applies to normal pulsars as well as MSPs is: What is the GeV emission mechanism?  As discussed in Section \ref{sec:emiss} and \ref{sec:LC}, CR, SR and IC have all been proposed.  A resolution of this issue could come from improvements in detected VHE emission components.  The maximum photon energy of the pulsed emission will provide a direct lower limit on the maximum energy of the accelerated particles.  The maximum photon energies of 2 TeV for Crab and around 7 TeV for Vela already requires particle energies of at least these energies.  This tends to disfavor SR and IC for the GeV emission, since the CR losses of particles with this high an energy can easily dominate.  Polarization observations could also distinguish between these different mechanisms since CR is predicted to have a higher degree of polarization than either SR or IC (Harding \& Kalapotharakos 2017).

There are many outstanding problems in the physics of emission from intra-binary shock accelerated particles.  The mechanism by which particles are accelerated in pulsar wind termination shocks has been an outstanding issue for decades primarily for pulsar wind nebulae.  It has now become a central question for the growing number of Black Widow and Redback binary systems that show emission from intra-binary shocks.  As discussed in Section \ref{sec:bin}, detection of a VHE emission component from IC can constrain the particle maximum energy.  Detecting both the VHE component and the cutoff of the SR spectrum, $E_{\rm SR} \propto B_s\gamma^2$, would give a measure of the magnetic field at the shock, $B_s$.  One might then constrain the $\sigma_w$ of the wind and distinguish between reconnection and DSA, if one of these mechanisms does not provide a consistent picture.  

I would like to thank Christo Venter for a careful reading of the manuscript and for many helpful suggestions, Zorawar Wadiasingh for comments and suggestions, and Matthew Baring for illuminating discussions.

\input{references-MSPemission}

\end{document}

%% file: references-MSPemission.tex
%
%
%